\documentclass[a4paper]{svproc}

\usepackage{url}

\usepackage{graphicx}
\usepackage{amsmath,amssymb}
\usepackage{booktabs}
\usepackage{algorithm}
\usepackage{algpseudocode}

\begin{document}
\mainmatter

\title{HQARRF: Hierarchical Q-learning and Force-aware Routing for\\
Multi-Charger Scheduling in Wireless Rechargeable Sensor Networks}
\titlerunning{HQARRF: Hierarchical Q-learning and Force-aware Routing for WRSNs}

\author{Liang-Ching Tao \and
Pi-Chung Wang\thanks{Corresponding author.}}
\authorrunning{L.-C. Tao and P.-C. Wang}
\tocauthor{Liang-Ching Tao, Pi-Chung Wang}

\institute{Department of Computer Science and Engineering,
National Chung Hsing University,\\
145 Xingda Rd., South District, Taichung 402, Taiwan\\
\email{liangkingtao@gmail.com}, \email{pcwang.tw@gmail.com}}

\maketitle

\begin{abstract}
Multi-charger scheduling in wireless rechargeable sensor networks must weigh sensor death
risk, charger energy, travel cost, return-to-base feasibility and inter-charger
coordination at once, and schedulers driven by local urgency alone duplicate service and
leave whole regions unattended. We present HQARRF, a two-level scheduler. Below, an
interpretable ARR-F score ranks candidate clusters through an attraction term for local
urgency, a repulsion term against charger crowding and a force bonus from nearby critical
sensors. Above, adaptive zones compress regional state into a deadline-based risk
estimate, and a gated Q-learning controller decides only whether to redirect service to a
high-risk, under-served zone. Over 27 parameter points HQARRF attains the highest mean
survival rate at 26, improving survival by 20.7 percentage points over the mean of five
baselines and 9.2 over the strongest baseline at each point. An ablation isolates the
upper level: its gain tracks how often the controller fires.
\end{abstract}

\section{Introduction}

Wireless rechargeable sensor networks (WRSNs) keep battery-powered sensor nodes alive
after deployment by delivering energy wirelessly rather than by replacing
batteries~\cite{xie2012making,he2013energyprovisioning,lu2016wirelesscharging}. A node
that forwards traffic for many others drains far faster than one reporting only its own
readings, so residual energy becomes uneven over time, and a node not recharged in time
opens a coverage hole or disconnects the subtree behind it.

Mobile chargers address this by touring the field and replenishing nodes in
place~\cite{zhang2018surveywpt,aziz2025wrsnsurvey}. Deploying several chargers raises
service capacity but creates a scheduling problem harder than route-length minimisation.
Each charger has a finite movement budget and must reserve enough energy to return to the
base station, so a locally attractive target is a poor choice if reaching it strands the
charger or starves another region; and when chargers evaluate targets independently,
several may converge on one neighbourhood while other high-risk regions receive no
service at all~\cite{beigel2014optimalscheduling,han2021mcca}.

Two observations motivate our design. Urgency is not purely local: several high-risk
regions can appear at once, and a scheduler that always picks the single most urgent
sensor allocates the fleet poorly. Yet the corrective signal needed to fix that is coarse
--- knowing \emph{which region} is under-served is enough, and no learned ranking over
sensor--charger pairs is required.

We therefore propose \textbf{HQARRF}, which separates these concerns into two levels. At
the lower level, an interpretable score named ARR-F ranks candidate clusters using an
electrostatic analogy: attraction drawn from local urgency, repulsion that discourages
charger crowding, and a force bonus contributed by nearby critical sensors. At the upper
level, adaptive zones aggregate regional state into a risk estimate, and a gated
Q-learning controller decides only \emph{whether} scheduling focus should be redirected
toward a high-risk, under-served zone. Confining learning to that one gated decision keeps
the table under two hundred entries in every configuration we ran and leaves the routing
behaviour explainable.

Our contributions are threefold. (i) We formulate ARR-F, a local routing score that
unifies urgency-driven attraction, anti-crowding repulsion and force-aware
critical-sensor selection under one electrostatic analogy, with target reservation
enforced separately as a hard constraint. (ii) We develop a zone-level risk estimator that converts per-sensor
time-to-death slack into a bounded deadline risk and aggregates the most endangered
sensors per zone, giving a regional view that candidate-cluster scoring alone cannot
provide. (iii) We evaluate against five baselines over 27 parameter points and, rather
than reporting the aggregate margin alone, use a four-step ablation to attribute it: the
variant that differs from HQARRF only by the absence of its upper level decays with
network size exactly as the baselines do, and the survival it recovers scales with how
often its controller fires.

\section{Related Work}

\paragraph{Charging scheduling.}
Early WRSN work optimised the tour of a single charger, establishing that the optimal
charging path is a shortest Hamiltonian cycle~\cite{xie2012making} and that energy
provisioning bounds network lifetime~\cite{he2013energyprovisioning}. On-demand schemes
let nodes request service once their energy falls below a
threshold~\cite{jiang2014ondemand,wang2014rechargingschedules}, which responds to actual
demand but leaves the threshold as a fixed global parameter; utility-based selection ranks
requests by a scalar value~\cite{ma2018chargingutility,chen2016chargeme}, and
partial-charging schemes trade per-node completeness for shorter
queues~\cite{wang2016mobiledatagathering}. Greedy service disciplines such as
nearest-job-next were characterised analytically by He et al.~\cite{he2015njnp}.

\paragraph{Multiple chargers.}
Once several chargers operate together, task allocation and coordination dominate.
Wei et al.~\cite{wei2019multimc} schedule multiple chargers under time windows;
Han et al. propose uneven cluster-based allocation~\cite{han2019ucmc} and explicit
multi-charger cooperation~\cite{han2021mcca}; dual-partition~\cite{jia2022dualpartition}
and hierarchical~\cite{madhja2016hierarchical} designs split the field to reduce
interference between chargers. These works establish that regional structure helps, but
the region assignment is generally static, so it cannot follow risk as it migrates
across the field.

\paragraph{Learning-based scheduling.}
Reinforcement learning has been applied to charging decisions in several
forms~\cite{jiang2022attentiondrl,jiang2023mddqn,li2024drlcoverage,mo2019energyaware}.
The common pattern is to learn the scheduling policy end to end, so the state space grows
with the number of sensor--charger pairs and the resulting routes are hard to explain.
The boundary we draw is one of scope rather than of technique: the learned component here
never selects a target, and the routing that does remains a closed-form score.

\section{System Model and Problem Statement}
\label{sec:model}

\begin{figure}[t]
\centering
\includegraphics[width=0.62\textwidth]{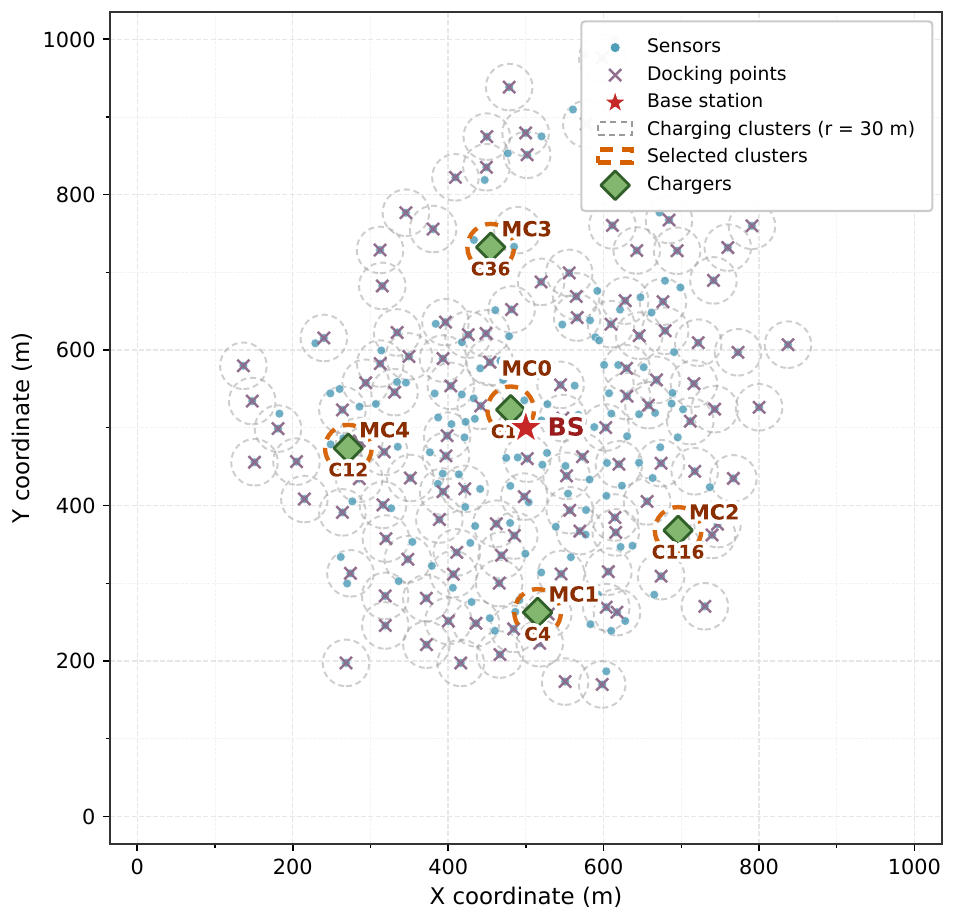}
\caption{A deployment at $t=0$: sensors, the candidate clusters produced by minimum
circle cover with their docking points, the base station, and the five chargers with the
clusters they have currently selected. Scheduling decisions are made over docking points
rather than over individual sensors.}
\label{fig:field}
\end{figure}

A set $V$ of static sensors is deployed over a square field of side $L$ with one base
station. Each sensor $i$ has capacity $E^{max}$ and residual energy $e_i(t)$ normalised to
$[0,1]$, and dies at the critical threshold. Sensors report over a multi-hop tree, so a
node's consumption rate $\rho_i$ covers its own sensing and transmission plus the relay
load of its subtree, and nodes closer to the base station drain fastest. Sensors are
grouped into candidate clusters $C$ by minimum circle cover, each cluster $c$ owning a
docking point $d_c$ at which a charger parks to serve the alive sensors within charging
radius $R_a$ (Fig.~\ref{fig:field}); deciding at cluster rather than sensor granularity
keeps the decision space manageable as the field grows~\cite{han2019ucmc}.

A fleet of $M$ mobile chargers, each with speed $v_m$, capacity $E_m^{max}$ and movement
power $P_m^{move}$, serves the field. Charging power decays with distance from the
docking point. An assignment of charger $m$ to cluster $c$ is \emph{feasible} only if
\begin{equation}
\label{eq:feasible}
E_m(t)\;\ge\;\frac{P_m^{move}}{v_m}\bigl(d_{m,c}+\lVert d_c-p_{BS}\rVert\bigr)
\;+\;E^{serve}_{c},
\end{equation}
where $E^{serve}_{c}$ is the energy the cluster's alive sensors need: the charger must be
able to reach $d_c$, serve it, and still return to the base station. Otherwise it returns
to recharge. The constraint is enforced on every assignment and is what makes a locally
attractive target sometimes inadmissible.

\subsubsection*{Problem.}
Let $A(T)$ be the number of sensors still alive at the end of a horizon $T$. The
scheduler chooses, at each decision epoch and for each idle charger, one feasible
cluster to serve (or the base station), so as to maximise $A(T)/|V|$ subject
to~(\ref{eq:feasible}) holding for every assignment and to each cluster being served by
at most one charger at a time. Travel distance and base-station recharge time are not
constrained but are reported, because a scheduler that maximises survival by spending
without limit is not a useful one; Section~\ref{sec:cost} states what HQARRF spends.

\section{The HQARRF Scheduler}

HQARRF is a two-level framework. The lower level answers \emph{which docking point is
worth serving now}; the upper level answers \emph{whether the fleet's focus should be
redirected to an under-served region}. Whichever level produces the target, the
assignment must still satisfy~(\ref{eq:feasible}). Algorithm~\ref{alg:hqarrf} gives one
decision epoch in full.

\begin{algorithm}[t]
\caption{HQARRF, one decision epoch (interval $\Delta$)}
\label{alg:hqarrf}
\begin{algorithmic}[1]
\Require clusters $C$ with docking points $d_c$; zones $Z$; chargers $\mathcal{M}$;
Q-table $Q$
\State recompute per-sensor deadline risk $u_i$ and zone risk $Risk_z^{smooth}$
\State $z^\star \gets \arg\max_z Risk_z^{smooth}$;\;
       form state $s_t=(z^\star,g_t,b_t)$
\State $a_t \gets \varepsilon$-greedy$(Q,s_t)$ \Comment{$a_t\in\{0,1\}$: redirect or not}
\If{$a_t=1$ \textbf{and} $z^\star$ passes the gate of~(\ref{eq:gate})}
  \State $C' \gets \{c \in C : d_c \in z^\star\}$ \Comment{intervention active}
\Else
  \State $C' \gets C$
\EndIf
\ForAll{idle chargers $m \in \mathcal{M}$}
  \State $C'_m \gets \{c \in C' : c$ unreserved and $(m,c)$ satisfies~(\ref{eq:feasible})$\}$
  \If{$C'_m = \emptyset$} \State send $m$ to the base station; \textbf{continue} \EndIf
  \State $c^\star \gets \arg\max_{c\in C'_m}\, S_c$ \Comment{ARR-F score,~(\ref{eq:arrf})}
  \State reserve $c^\star$ for $m$ and dispatch
\EndFor
\State observe reward $r_t$; $Q(s_t,a_t) \gets Q(s_t,a_t) +
       \alpha\bigl[r_t+\gamma\max_{a}Q(s_{t+1},a)-Q(s_t,a_t)\bigr]$
\end{algorithmic}
\end{algorithm}

\subsection{ARR-F Local Routing Score}

For candidate cluster $c$, the local score combines three terms,
\begin{equation}
\label{eq:arrf}
S_c = \widetilde{A}_c - \beta\,\widetilde{\Phi}_c + \lambda_f F_c ,
\qquad c^\star = \arg\max_{c\in C'_m} S_c ,
\end{equation}
where $\widetilde{A}_c$ and $\widetilde{\Phi}_c$ are the attraction and repulsion terms
below, each min--max normalised over the candidate set of the current epoch so that the
two are commensurable, and $\beta,\lambda_f$ weight repulsion and the force bonus.
Reservation does not appear in~(\ref{eq:arrf}): a cluster already claimed by another
charger is removed from $C'_m$ outright. Repulsion and reservation are therefore distinct
mechanisms, and the ablation of Section~\ref{sec:ablation} isolates repulsion by setting
$\beta=0$ while leaving reservation in place.

\paragraph{Attraction.}
Cluster urgency is a weighted sum of the cluster's normalised energy deficit, mean
consumption rate, criticality and alive count, together with an asymmetric term in the
blended residual-energy ratio that is positive below a cutoff of $0.8$ and sharply
negative above it. That sum is min--max normalised over the candidates to
$\widehat{U}_c\in[0,1]$ and compressed into a scalar charge
$Q_c = 1-2\widehat{U}_c \in[-1,1]$. Attraction then follows an inverse-square law between
the charger at $p_m(t)$ and the docking point,
\begin{equation}
\label{eq:attraction}
A_c = -\frac{k_a q_m Q_c}{d_{m,c}^{2}+\epsilon_a^{2}} ,
\qquad d_{m,c}=\|p_m(t)-d_c\| ,
\end{equation}
with charge gain $k_a$, charger charge $q_m$ and a softening radius
$\epsilon_a=\max(\tfrac12 R_a,1)$ that keeps the score finite at contact. The sign
convention carries the intent: an urgent cluster has $\widehat{U}_c\to1$, hence
$Q_c\to-1$, which the leading minus sign turns into strong attraction, while a
well-charged cluster has $Q_c\to+1$ and becomes mildly repulsive, suppressing repeated
visits without any explicit exclusion rule.

\paragraph{Repulsion.}
$\Phi_c$ is a Coulomb-like penalty accumulated from the current positions of the other
chargers,
\begin{equation}
\label{eq:repulsion}
\Phi_c=\sum_{m'\neq m}\frac{1}{\lVert p_{m'}(t)-d_c\rVert^{2}+\epsilon_r^{2}} ,
\qquad \epsilon_r=\max(\tfrac12 R_a,1) .
\end{equation}
Independent evaluation of the same urgency signal makes crowding the default failure mode
even when reservation prevents outright duplication, since chargers converge on adjacent
docking points around one hot region. Repulsion makes a cluster progressively less
attractive as other chargers approach it, so the fleet spreads without centralised
assignment.

\paragraph{Force bonus.}
Let $\mathcal{T}_{K_f}(c)$ be the set of at most $K_f$ alive sensors whose residual energy
ratio does not exceed a force threshold $\theta_f$ and that lie within radius $R_f$ of
$d_c$. Then
\begin{equation}
\label{eq:force}
F_c=\log\!\left(1+\sum_{i\in\mathcal{T}_{K_f}(c)}
\frac{1-e_i(t)}{\|p_i-d_c\|^{2}+\epsilon_f}\right),
\end{equation}
where $\epsilon_f$ is a numerical stabiliser. A docking point thus inherits value from
critical sensors that sit near it without belonging to it --- which matters at cluster
boundaries, where the nearest endangered sensor is often just outside the cluster that
would serve it --- while the logarithm stops a dense pocket of critical nodes dominating
$S_c$.

\subsection{Adaptive Zones and Zone-Level Risk}

Zones aggregate regional state, and are coarser than clusters: in our runs the field
holds $124$--$633$ candidate clusters but only $7$--$11$ zones. Their number is derived
once per deployment rather than tuned. Write $n_{eff}=(\sum_i \rho_i)^2/\sum_i \rho_i^2$
for the effective network size under per-node consumption rates $\rho_i$, and
$\lambda_d = n_{eff}\pi R_a^2/L^2$ for the coverage density. A node-driven requirement
grows with both,
\begin{equation}
\label{eq:znode}
z_{node}=\bigl\lceil (1+\ln n_{eff})\max(1,\lambda_d) \bigr\rceil ,
\end{equation}
a space-driven requirement $z_{space}=\lceil L^{2}/(\pi r_{resp}^{2})\rceil$ follows from
the distance $r_{resp}$ a charger can cover before sensors start dying, a learnability cap
$z_{learn}$ keeps the Q-table small enough to fill within the horizon, and
\begin{equation}
\label{eq:zonecount}
|Z|=\max\Big(1,\ \min\big(|C|,\ z_{learn},\ \max(2,\ z_{node},\ z_{space})\big)\Big),
\end{equation}
with boundaries then obtained by $k$-means over sensor positions. In every configuration
we ran the binding term was $z_{node}$, giving $|Z|=7$ to $11$; $z_{space}$ and
$z_{learn}$ acted only as guards.

Risk is built from deadlines rather than from energy alone. For sensor $i$, let $TTD_i$
be its time to death at the current consumption rate and
$ETA_i=\min_m\|p_i-p_m(t)\|/v_m$ the earliest arrival of any charger; the slack
$slack_i=TTD_i-ETA_i$ is mapped to a bounded per-sensor risk
\begin{equation}
\label{eq:urisk}
u_i=\begin{cases}
\exp(-slack_i/\tau), & slack_i \ge 0,\\[2pt]
1+\delta, & slack_i < 0,
\end{cases}
\end{equation}
bounded above by $1+\delta$, so that a sensor no charger can reach in time saturates the
scale instead of competing on the same continuum with sensors that are merely far away;
$\tau$ sets how fast comfortable slack decays and $\delta$ is the missed-deadline
bonus. When the adaptive threshold of Section~4.3 is active it multiplies $u_i$ by up to
$1.5$ for sensors in locally stressed neighbourhoods. Zone risk aggregates only the most
endangered members,
\begin{equation}
\label{eq:zonerisk}
Risk_z^{raw}=\frac{1}{\min\left(K_z,\left|V_z^{alive}\right|\right)}
\sum_{i\in\mathcal{T}_{K_z}(z)}u_i ,
\end{equation}
where $\mathcal{T}_{K_z}(z)$ holds the $K_z$ alive sensors in $z$ with the largest $u_i$.
Averaging over the top-$K_z$ rather than the whole zone stops a small pocket of genuinely
endangered sensors being diluted by healthy neighbours, and normalising by
$\min(K_z,|V_z^{alive}|)$ stops a nearly depleted zone being assigned an artificially low
risk merely because too few nodes survive to fill the set. The value used downstream is
smoothed over the $k$ nearest zones,
$Risk_z^{smooth}=(1-\lambda_s)Risk_z^{raw}+\lambda_s\overline{Risk}_{\mathcal{N}(z)}$, so
that zone boundaries do not act as hard discontinuities.

\subsection{Gated Q-learning Intervention}

The upper level observes a compact state and encodes it as a single index,
\begin{equation}
\label{eq:state}
s_t=(z_t^\star, g_t, b_t), \qquad
z_t^\star=\arg\max_{z} Risk_z^{smooth}, \qquad
\mathrm{id}(s_t)=9z_t^\star+3g_t+b_t ,
\end{equation}
where $g_t,b_t\in\{0,1,2\}$ are the service capacity of $z_t^\star$ (how many chargers
could reach it feasibly) and its coverage ratio (the share of its critical sensors
already inside some charger's service radius), each discretised into three levels. The
action is binary: intervene or not. Tabular Q-learning~\cite{watkins1992qlearning} is
sufficient at this granularity: with $|Z|\le 11$ the table has at most $9\cdot 11\cdot 2 =
198$ entries, of which between $22$ and $186$ were ever updated across our runs, so the
learned policy can be printed and inspected.

Intervention is additionally gated. A zone is eligible only if
\begin{equation}
\label{eq:gate}
Risk_{z^\star}^{smooth}\ge\theta_{risk},\qquad g_t>0,\qquad
\text{coverage}(z^\star)<\theta_{cov},
\end{equation}
so the controller cannot fire on a low-risk zone, a zone no charger can reach, or one
that is already sufficiently covered; across the 27 parameter points $92\%$ of the
epochs in which the policy chose to intervene passed this gate. When intervention is
active, ARR-F is restricted to candidate clusters inside $z_t^\star$; otherwise routing
proceeds unrestricted. Learning is driven by a reward combining the change in alive
nodes, requests served, critical-node count and delivered energy, against penalties on
newly dead nodes, extra movement and the act of intervening.

Two smaller upper-level components are enabled alongside the controller, and
Section~\ref{sec:ablation} therefore measures all three as one step: a target selector
that ranks by time-to-death rather than urgency alone, and a soft adaptive threshold that
raises a sensor's effective urgency in locally stressed neighbourhoods.

\section{Evaluation}
\label{sec:eval}

\subsection{Setup}

Five scenario groups vary sensor count, field size, sensor battery capacity, charger speed
and consumption rate, giving 27 parameter points. Each group passes through the same
default configuration, so the 27 points cover 24 distinct ones; collapsing the repeats
moves the aggregate margins below by less than $0.3$ percentage points. Every point is
run on five independent network instances from one fixed base seed, and every method sees
the same five instances, so every comparison in this campaign is paired. Table~\ref{tab:params} lists the
parameters. The primary metric is sensor survival rate at the end of the horizon; travel
distance, delivered energy, base-station recharge time and per-charger load balance are
reported alongside it.

\begin{table}[t]
\centering
\caption{Simulation parameters. Swept values are given as ranges with the default, used
when another parameter is swept, in bold.}
\label{tab:params}
\begin{tabular}{llll}
\toprule
Field side $L$ & 1000--3000\,m (\textbf{1000}) & Charger capacity $E_m^{max}$ & $10^4$\,J \\
Sensors $|V|$ & 300--500 (\textbf{250})$^{\dagger}$ & Charger speed $v_m$ & 1--10\,m/s (\textbf{5}) \\
Base station & field centre & Movement power $P_m^{move}$ & 0.1\,W \\
Sensor battery $E^{max}$ & 100--400\,J (\textbf{150}) & Charging radius $R_a$ & 30\,m \\
Consumption $\rho_i$ & 0.5--1.5\,mJ/s (\textbf{1.0}) & Charging power & 10\,W \\
Sensing range & 40\,m & Charging efficiency & 0.9 \\
Comm.\ range & 80\,m & Chargers $M$ & \textbf{5} (2--7 in \S\ref{sec:fleet}) \\
\midrule
Repulsion weight $\beta$ & 0.2 & Risk top-$K_z$ & 5 \\
Force weight $\lambda_f$ & 1.0 & Risk decay $\tau$ & 8000\,s \\
Force radius $R_f$ & 150\,m & Missed-deadline $\delta$ & 0.5 \\
Force cap $K_f$ & 10 & Risk gate $\theta_{risk}$ & 0.45 \\
Force threshold $\theta_f$ & 0.3 & Coverage gate $\theta_{cov}$ & 0.8 \\
Decision interval $\Delta$ & 30\,s & Smoothing $\lambda_s$ ($k{=}3$) & 0.3 \\
Q-learning $\alpha,\gamma$ & 0.1, 0.9 & $\varepsilon$ (start/min/decay) & 0.3 / 0.05 / 0.995 \\
Horizon $T$ & $10^5$\,s & Runs per point & 5 \\
\bottomrule
\end{tabular}\\[2pt]
{\footnotesize $^{\dagger}$The field-size sweep scales $|V|$ with area, from 250 at
1000\,m to 750 at 3000\,m, so node density stays constant; the sensor-count sweep varies
$|V|$ from 300 to 500 at $L=1000$\,m.}
\end{table}

We compare against five baselines spanning the main design families: a genetic algorithm
with 2-OPT tour improvement under time windows~\cite{wei2019multimc}, multi-charger
cooperation (MCCA)~\cite{han2021mcca}, uneven cluster-based charging
(UCMC)~\cite{han2019ucmc}, a $k$-way earliest-deadline-first heuristic and
nearest-job-next. The first three follow their published designs. The last two have no
single canonical multi-charger formulation, so we implement them as family
representatives rather than as reproductions of a specific paper: K-EDF sorts pending
requests by residual lifetime in the classical earliest-deadline-first
order~\cite{liu1973edf} and assigns the top $k$ to available chargers by minimum total
travel; NJNP sends each charger to its nearest pending request, following the greedy
nearest-job discipline analysed for on-demand charging by He et al.~\cite{he2015njnp} but
without their preemption rule. All five share the same simulation core --- energy model,
distance-decay charging model, charging radius and charger speed --- and differ only in
target selection and scheduling policy. Baselines whose original formulations do not model
the charger's own energy budget were extended with the same return-to-base
check~(\ref{eq:feasible}) that HQARRF must satisfy.

\subsection{Survival Rate}

Across the 27 parameter points, HQARRF attains the highest mean survival rate at 26.
Relative to the mean of the five baselines the improvement is 20.7 percentage points;
relative to the strongest baseline \emph{at each individual point} it is 9.2 points.
The margin is not uniform across the sweeps, and in the lightest-load regimes the leading
methods sit within each other's run-to-run variation --- including the single point HQARRF
does not lead, where the gap is $0.16$ points. The figures carry error bars throughout.

\begin{figure}[t]
\centering
\includegraphics[width=\textwidth]{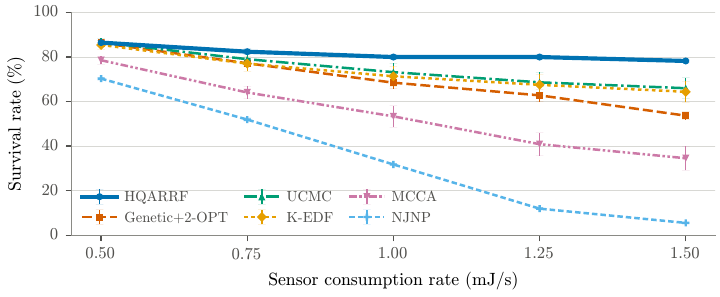}
\caption{Survival rate versus sensor consumption rate. Whiskers are $\pm1$ standard error
over the five runs at each point. At 0.5\,mJ/s the leading methods lie within each
other's error bars; as workload rises HQARRF degrades gracefully while the baselines fall
away.}
\label{fig:consumption}
\end{figure}

The shape of the margin is informative: 26.1 points in the sensor-count group, 3.7 in the
battery-capacity group. Fig.~\ref{fig:consumption} makes the pattern concrete along the
workload axis. There is little for zone-level intervention to contribute
while charging pressure is low, but as consumption rises the methods separate
monotonically: HQARRF degrades gently from 86.5\% to 78.2\%, the strongest baseline falls
to 66.0\% and the weakest collapses to 5.5\%, opening the margin from $-0.2$ to $+12.2$
points without reversal.

\begin{figure}[t]
\centering
\includegraphics[width=\textwidth]{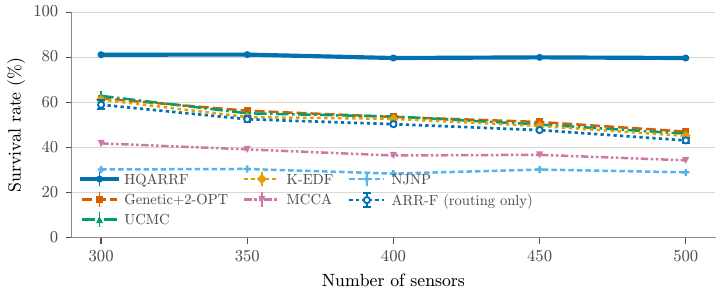}
\caption{Survival rate versus network size. HQARRF holds near 80\% while every other
method decays. ARR-F is HQARRF with its upper level removed and the routing score
unchanged; it decays by 15.8 points, placing the flatness in the upper level rather than
the routing score.}
\label{fig:sensors}
\end{figure}

Fig.~\ref{fig:sensors} shows the same ordering along network size, and adds the evidence
that makes the reading causal rather than suggestive. HQARRF loses only $1.5$ points
between 300 and 500 sensors while the three strongest baselines lose $14.7$ to $16.8$ ---
which by itself would not say which part of HQARRF is responsible. The dotted line is
ARR-F, which is HQARRF with its upper level removed and the routing score unchanged: it
decays by $15.8$ points, indistinguishably from the baselines, placing the flatness with
the regional view rather than the local score. NJNP is also nearly flat, at $-1.3$ points,
but flat at roughly 30\% throughout, having already collapsed at the smallest network.

\subsection{Component Ablation}
\label{sec:ablation}

Four variants isolate the contributions. AR uses attraction with hard reservation only
($\beta=0$); ARR adds the repulsion term; ARR-F adds the force bonus; HQARRF adds the
gated zone-level controller together with the two upper-level components named at the end
of Section~4.3. Averaged over the 27 points, survival rises $48.8 \to 63.9 \to 69.8 \to
81.6$\%, and each step is an improvement at every one of the 27 points individually
(mean increments $+15.1$, $+5.9$ and $+11.8$ points).

Travel does not behave as a spreading argument would predict. Each added component
\emph{reduces} average travel rather than raising it, from $2240$\,km for AR to $1980$,
$1950$ and $1760$\,km; the reductions from repulsion and from the controller hold at all
27 points, while the force bonus is travel-neutral (lower at 10 of 27). Better target
choice removes wasted trips: a charger not sent to a cluster another charger is about to
serve, or to a region that will be covered anyway, does not pay the round trip. Survival
and travel improve together here; the trade-off HQARRF does pay appears in
Section~\ref{sec:cost}.

\begin{figure}[t]
\centering
\includegraphics[width=\textwidth]{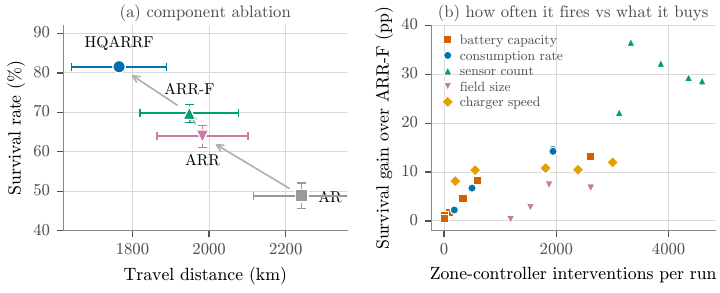}
\caption{(a) The four variants on the survival--travel plane, averaged over the 27
parameter points; whiskers are $\pm1$ standard error \emph{across those points}, so they
show how much the scenarios differ, not run-to-run noise. The survival ordering holds at
all 27 points individually. (b) For each point, how often the zone controller engaged
against the survival it recovered relative to ARR-F. The relationship is monotone within
the battery and consumption sweeps and near-monotone across charger speed; the two sweeps
that change the field itself sit at the high end of both axes.}
\label{fig:ablation}
\end{figure}

The last step carries the paper's claim, and the ablation supports it more directly than
an increment can. The controller is dormant by design --- the gate of~(\ref{eq:gate})
keeps it idle unless a zone is both at risk and under-served --- so what it recovers over
ARR-F should scale with how often it actually engages. It does: across the 27 points the
survival it adds rises with its intervention count (Fig.~\ref{fig:ablation}(b),
$r=0.84$). It is monotone in the battery-capacity and consumption-rate sweeps and
near-monotone across charger speed --- the three that vary charging pressure at a fixed
topology. The two sweeps that change the field itself instead sit at the top of both
axes: in the largest networks the controller engages almost constantly and recovers up to
$36.6$ points, the most it contributes anywhere. Because
ARR-F differs from HQARRF only in that upper level, the gain tracks the mechanism rather
than the routing score, and it explains the shape of the margin reported above: the
advantage is largest exactly where sensor death would otherwise be irreversible.

\subsection{Fleet Size}
\label{sec:fleet}

\begin{figure}[t]
\centering
\includegraphics[width=\textwidth]{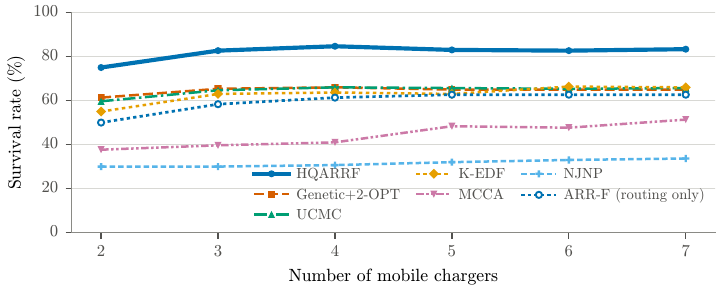}
\caption{Survival rate versus fleet size, at 300 sensors. Every method saturates by four
chargers without the methods converging.}
\label{fig:chargers}
\end{figure}

A separate sweep varies the fleet from two to seven chargers at 300 sensors
(Fig.~\ref{fig:chargers}).
Survival rises steeply from two to four and is essentially flat thereafter for every
method, so the fleet saturates --- but the methods do not converge as they saturate.
HQARRF leads at every fleet size tested, by 13.7 points at two chargers and by 16.3 to
18.7 from three upward, with ARR-F again tracking the baselines rather than HQARRF. What
this sweep establishes is the separation between methods; the small variations along each
curve are not meant to be read.

\subsection{What the Gain Costs}
\label{sec:cost}

The survival advantage is not free, but the cost is not where a route-length argument
would put it. Averaged over the 27 points HQARRF travels $1760$\,km, the lowest of the
five non-route-optimising methods (UCMC $1820$, MCCA $1830$, K-EDF $1900$, NJNP
$2440$\,km). Only the genetic baseline with 2-OPT travels less, and by a wide margin: at
$367$\,km it covers $4.8$ times less ground, which is what a scheduler optimising tours
for their own sake should do, and it pays $13.8$ points of survival for it.

The real cost is service throughput. HQARRF delivers the most energy of any method
($27.6$\,kJ against $21.6$\,kJ for the genetic baseline) and so must refuel most often:
base-station recharge time averages $76.0\%$ of the horizon against $64.3\%$ for UCMC and
$61.2\%$ for the genetic baseline, the highest of the six at 20 of the 27 points. A
charger that is refuelling is not serving, so this is the first thing a deployment with
slow base-station charging should check. Load balance is unremarkable: the coefficient of
variation of travel across the fleet averages $0.12$, matching UCMC and far below the
genetic baseline's $0.73$, though above the near-uniform $0.01$--$0.04$ of the round-robin
heuristics.

The case for HQARRF is therefore not that it dominates every axis, but that the
throughput it sustains buys a survival improvement the cheaper methods do not achieve,
and that the exchange is favourable precisely in the high-pressure regimes where sensor
death would otherwise be irreversible. Where charging pressure is low the controller
barely fires, the margin falls inside the noise, and a simpler scheduler is the better
choice.

\subsection{Reproducibility}

Every number above comes from one simulation campaign, exported to flat CSV at three
granularities --- per run, per point and per charger --- together with the per-point
intervention counts extracted from the run summaries. One script regenerates all four
result figures from those files, so each plotted value is traceable to a stored run
rather than read back off an image. The simulator writes a snapshot of its full
configuration with each campaign, and Table~\ref{tab:params} is taken from the snapshot
of the campaign reported here rather than from the current defaults.

\section{Conclusion}

We presented HQARRF, a two-level scheduler for multi-charger WRSNs that keeps routing
interpretable and confines learning to a gated, zone-level intervention decision. Across
27 parameter points it leads at 26, improving survival by 20.7 percentage points over the
baseline mean and 9.2 points over the strongest baseline at each point. The ablation does
more than rank the components: the variant without the upper level decays with network
size exactly as the baselines do, and the survival it recovers scales with how often the
controller fires, which locates the gain in the regional view rather than the local
score. The framework should be read as risk-aware resource allocation rather than
route-distance minimisation --- it travels less than the comparable heuristics, but it
spends three quarters of the horizon refuelling to sustain the energy it delivers, and
that is the cost to weigh.

Three directions follow. The first is validation outside simulation, through
hardware-in-the-loop testing or a small physical deployment, which would show how the
charging model and the return-to-base design behave under real localisation and terrain
error. The second is automatic parameter adaptation: the risk threshold, coverage
threshold and zone count are fixed per scenario here, and could instead be mapped from
observable scenario features. The third follows from the cost analysis --- since
base-station refuelling rather than travel is what limits the method, charger-to-charger
transfer or additional depots are the natural next lever.

\subsubsection*{Acknowledgements.}
This work was supported by the National Science and Technology Council, Taiwan, under
Grant No.\ NSTC 114-2221-E-005-043-MY3. Parts of this work appeared in the first author's
master's thesis at National Chung Hsing University.

\bibliographystyle{spmpsci}
\bibliography{refs}

\end{document}